\documentclass[lettersize,journal]{IEEEtran}
\usepackage{setspace}
\usepackage{amsmath,amsfonts}
\usepackage{algorithmic}
\usepackage{algorithm}
\usepackage{array}
\usepackage[caption=false,font=normalsize,labelfont=small,textfont=small]{subfig}
\usepackage{textcomp}
\usepackage{stfloats}
\usepackage{url}
\usepackage{color}
\usepackage[justification=centering]{caption}
\usepackage{verbatim}
\usepackage{bbding}
\usepackage{threeparttable}
\usepackage{graphicx}
\usepackage{cite}
\usepackage[normalem]{ulem}
\useunder{\uline}{\ul}{}
\begin{document}

\title{Stacked Intelligent Metasurface-Aided\\MIMO Transceiver Design}


\author{Jiancheng An,~\emph{Member, IEEE}, Chau Yuen,~\emph{Fellow, IEEE}, Chao Xu,~\emph{Senior Member, IEEE},\\Hongbin Li,~\emph{Fellow, IEEE}, Derrick Wing Kwan Ng,~\emph{Fellow, IEEE}, Marco Di Renzo,~\emph{Fellow, IEEE},\\M\'erouane Debbah,~\emph{Fellow, IEEE}, and Lajos Hanzo,~\emph{Life Fellow, IEEE}
\thanks{J. An and C. Yuen are with the School of Electrical and Electronics Engineering, Nanyang Technological University, Singapore 639798 (e-mail: jiancheng.an@ntu.edu.sg, chau.yuen@ntu.edu.sg). C. Xu and L. Hanzo are with the School of Electronics and Computer Science, University of Southampton, Southampton SO17 1BJ, UK (e-mail: cx1g08@ecs.soton.ac.uk; lh@ecs.soton.ac.uk). H. Li is with the Department of Electrical and Computer Engineering, Stevens Institute of Technology, Hoboken, NJ 07030 USA (e-mail: hli@stevens.edu). D. W. K. Ng is with the School of Electrical Engineering and Telecommunications, University of New South Wales, Sydney, NSW 2052, Australia (e-mail: w.k.ng@unsw.edu.au). M. Di Renzo is with Universit\'e Paris-Saclay, CNRS, CentraleSup\'elec, Laboratoire des Signaux et Syst\`emes, 3 Rue Joliot-Curie, 91192 Gif-sur-Yvette, France (e-mail: marco.di-renzo@universite-paris-saclay.fr). M. Debbah is with the Center for 6G Technology, Khalifa University of Science and Technology, P O Box 127788, Abu Dhabi, United Arab Emirates (e-mail: merouane.debbah@ku.ac.ae).}}
\maketitle

\begin{abstract}
Next-generation wireless networks are expected to utilize the limited radio frequency (RF) resources more efficiently with the aid of intelligent transceivers. To this end, we propose a promising transceiver architecture relying on stacked intelligent metasurfaces (SIM). An SIM is constructed by stacking an array of programmable metasurface layers, where each layer consists of a massive number of low-cost passive meta-atoms that individually manipulate the electromagnetic (EM) waves. By appropriately configuring the passive meta-atoms, an SIM is capable of accomplishing advanced computation and signal processing tasks, such as multiple-input multiple-output (MIMO) precoding/combining, multi-user interference mitigation, and radar sensing, as the EM wave propagates through the multiple layers of the metasurface, which effectively reduces both the RF-related energy consumption and processing delay. Inspired by this, we provide an overview of the SIM-aided MIMO transceiver design, which encompasses its hardware architecture and its potential benefits over state-of-the-art solutions. Furthermore, we discuss promising application scenarios and identify the open research challenges associated with the design of advanced SIM architectures for next-generation wireless networks. Finally, numerical results are provided for quantifying the benefits of wave-based signal processing in wireless systems.
\end{abstract}

\section{Introduction}
Massive multiple-input multiple-output (mMIMO) techniques have become one of the most innovative technical enablers towards ubiquitous mobile connectivity. Although mMIMOs attain tremendous throughput gains, they inevitably increase the associated hardware complexity and energy consumption \cite{CM_2014_Alkhateeb_MIMO}. Additionally, next-generation wireless networks are expected to intrinsically integrate communication, sensing, computing, and control capabilities, while improving the throughput, latency, and connectivity \cite{CST_2022_Zhang_Enabling}. Hence, it is imperative to develop radically new techniques for next-generation wireless networks.

Recent contributions have demonstrated the groundbreaking potential of metasurfaces in realizing these ambitious goals \cite{JSAC_2020_Di_Smart, TCOM_2022_An_Low}. Generally speaking, a programmable metasurface is an ultra-thin planar structure engineered by compactly arranging a large number of low-cost electrically controllable metamaterial particles, each of which is capable of independently imposing an adjustable attenuation and/or phase shift onto the electromagnetic (EM) waves impinging on it \cite{TCOM_2022_Liu_Compact}. An example of a programmable metasurface is the so-called reconfigurable intelligent surface (RIS), which, thanks to a field-programmable gate array (FPGA), can beneficially customize the wireless propagation environment \cite{JSAC_2020_Di_Smart}. By jointly optimizing the RF-powered active beamforming at the transmitter and the RF-free nearly passive beamforming at the RIS, several research contributions have demonstrated the capability of an RIS to improve the performance of existing wireless networks at a modest cost. However, the joint transceiver-RIS beamforming optimization process may have to be executed frequently because of the dynamic nature of the wireless environments, thus often resulting in an increased signal processing complexity \cite{TCOM_2022_An_Low}. In addition, the distributed deployment of several RISs may also increase the control signaling overhead and may complicate the beam management protocol in the medium access control (MAC) layer. Furthermore, a single-layer nearly passive RIS architecture lacks the capability of implementing advanced MIMO functionalities due to intrinsic hardware limitations and the severe path-loss attenuation of the transmitter-RIS-receiver path if the location of the RIS is not appropriately optimized \cite{TCOM_2022_An_Low}. All in all, deploying an RIS in the vicinity of the transceivers has been demonstrated to be the most energy-efficient solution \cite{JSAC_2020_Di_Smart}.

Motivated by the rapid development of metasurface design, the innovative concept of holographic MIMO is envisioned to ease the implementation of mMIMO by integrating an electromagnetically active metasurface into the wireless transceiver architecture \cite{COM_2021_Dardari_Holographic}. By integrating a virtually continuum number of tiny radiating elements into a compact surface, a spatially near-continuous aperture can be created to form ``pencil'' beams with extremely low sidelobe leakage. Without breakthroughs in metasurface design, the implementation of holographic MIMO would suffer from excessive energy consumption and hardware cost due to the use of an increased number of active components. In a single-layer metasurface architecture, in addition, the number of tunable amplitude/phase associated with each meta-atom is limited by practical hardware constraints \cite{COM_2021_Dardari_Holographic}. In summary, while existing holographic MIMOs have an ever-denser array layout, energy-efficient transceiver techniques are urgently needed.

\begin{figure*}[!t]
\centering
\includegraphics[width=17cm]{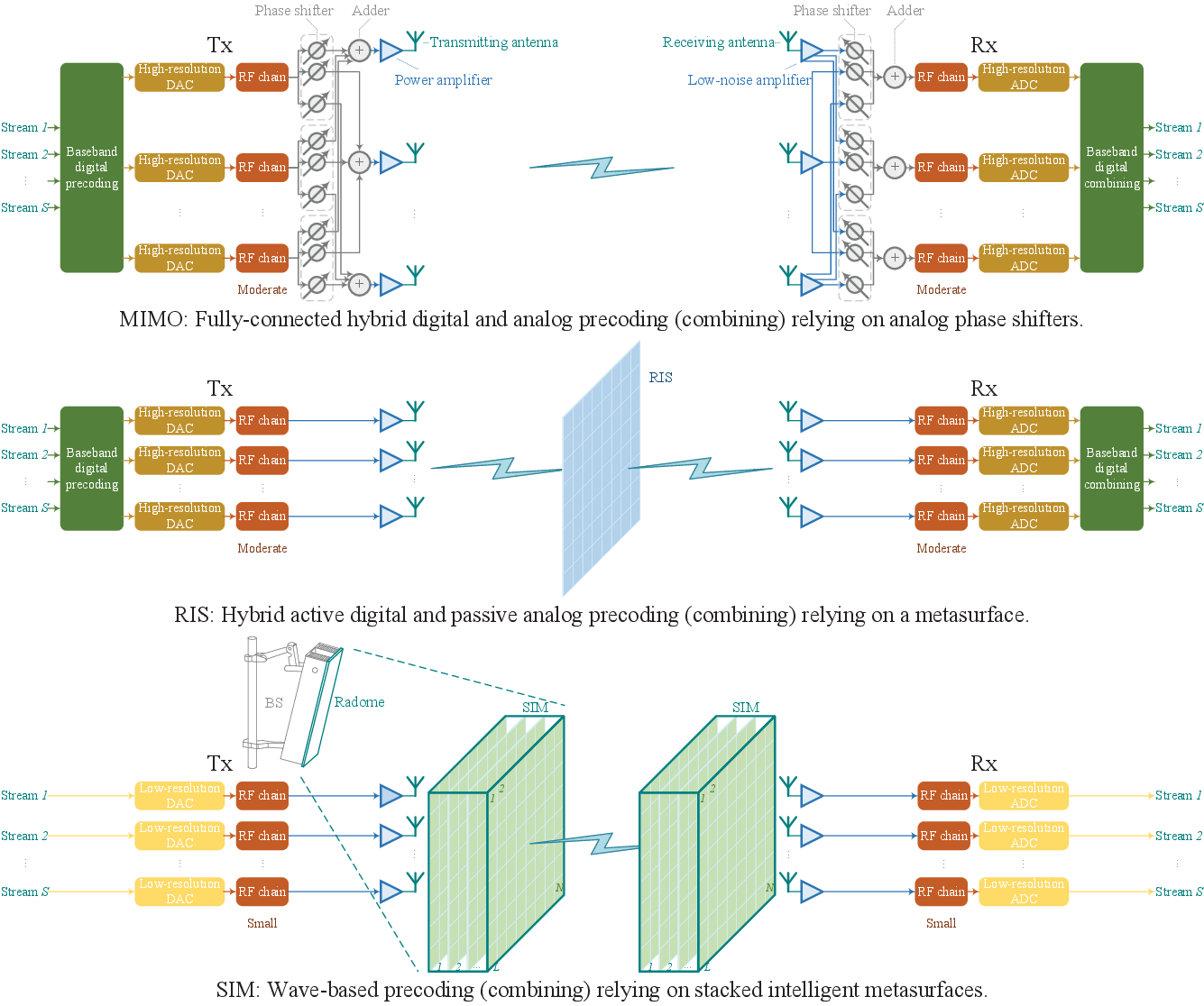}
\caption{Contrasting an SIM-aided MIMO transceiver to its conventional counterparts.}\vspace{-0.5cm}
\label{fig_1}
\end{figure*}
Recently, a radical signal processing paradigm aided by metasurfaces has attracted significant research attention \cite{NE_2022_Liu_A, Science_2018_Lin_All, arXiv_2023_An_Stacked, ChemPhysMater_2022_Yu_Electromagnetic}. Specifically, \emph{Liu et al.} \cite{NE_2022_Liu_A} constructed a programmable diffractive deep neural network architecture by utilizing a multilayer metasurface array, where each meta-atom acts as a reprogrammable artificial neuron. This new device is capable of executing various complex signal processing tasks, such as image classification, directly in the native EM wave domain. Motivated by this, \emph{An et al.} proposed a stacked intelligent metasurface (SIM)-aided MIMO transceiver \cite{JSAC_2022_An_Stacked, ICC_2023_An_Stacked}. By stacking an array of multiple programmable nearly passive metasurfaces, an SIM having a structure similar to an artificial neural network (ANN) offers substantial signal processing capabilities. Most remarkably, the signal propagation in an SIM is at the speed of light. As such, an SIM ultimately facilitates the implementation of MIMO transceivers with advanced transmit precoding and receive combining capabilities in the wave domain, while substantially reducing the RF energy consumption and hardware cost.

An SIM-aided MIMO transceiver architecture is profoundly distinct from its conventional counterparts. This article presents a systematic overview of SIM, covering the hardware architecture, potential technological advantages, and promising applications in wireless networks. Furthermore, we discuss open research challenges and opportunities, such as efficient SIM designs, the inter-layer transmission model, channel state information (CSI) acquisition, and wave-based beamforming (WBF) design. Finally, numerical results are provided for characterizing the performance offered by an SIM in typical sensing and communication scenarios.

\section{Hardware Architecture and Benefits of SIM}
In this section, we present the general hardware architecture of SIM. Then, we elaborate on the benefits of SIM-aided MIMO transceivers.
\subsection{Hardware Architecture}
As shown at the bottom of Fig. \ref{fig_1}, an SIM is physically fabricated by stacking an array of metasurfaces. Each metasurface layer comprises a large number of low-cost, passive meta-atoms that are electronically tunable by relying on off-the-shelf electronic components like PIN diodes or reconfigurable metamaterials such as liquid crystals or graphene \cite{JSAC_2020_Di_Smart}. With the aid of a smart controller, e.g., a customized FPGA, the EM responses of each meta-atom -- including both its amplitude and phase shift -- can be individually adjusted by controlling the bias voltage of the associated control circuit. As a result, the SIM becomes capable of dynamically adapting its EM properties on demand. In accordance with the Huygens–Fresnel principle \cite{Science_2018_Lin_All}, the transmit EM wave that goes through a meta-atom in each layer acts as a secondary point source and illuminates all the meta-atoms in the subsequent layer. Furthermore, all EM waves impinging on a meta-atom in a metasurface layer are superimposed, acting as a wave incident onto this meta-atom. By appropriately designing the complex-valued transmission coefficient of the meta-atoms in each metasurface, an SIM becomes capable of hierarchically manipulating the energy distribution of the EM waves going through it, and accomplishing various advanced signal processing tasks, such as MIMO precoding and receive combining operations, directly in the EM wave domain \cite{NE_2022_Liu_A}. In practice, an SIM can serve as a radome of a typical base station (BS) (see Fig. \ref{fig_1}) to effectively unleash its benefits.

\begin{table*}[!t]
\centering
\footnotesize
\renewcommand\arraystretch{1.5}
\caption{Contrasting SIM to relevant concepts and technologies.}
\label{tab1}
\begin{tabular}{lllllll}
\hline
\multicolumn{7}{c}{Comparison of SIM with existing concepts}                                     \\ \hline
\multicolumn{1}{l|}{Existing concepts}                   & \multicolumn{3}{l|}{Features}                                                                                                                                                                                                                     & \multicolumn{3}{l}{SIM's fascinating features}                                                                                                                                                                                                                                               \\ \hline
\multicolumn{1}{l|}{RIS}                                 & \multicolumn{3}{l|}{\begin{tabular}[c]{@{}l@{}}A single-layer metasurface for reconfiguring\\the wireless propagation environments.\end{tabular}}                                                                                                             & \multicolumn{3}{l}{\begin{tabular}[c]{@{}l@{}}$\star$ A  multi-layer metasurface architecture for enabling\\more powerful computing capability.\end{tabular}}                                                                                                                                        \\ \hline
\multicolumn{1}{l|}{ANN}                                 & \multicolumn{3}{l|}{\begin{tabular}[c]{@{}l@{}}A multi-layer neural network to accomplish\\complex computation tasks.\end{tabular}}                                                                                                                             & \multicolumn{3}{l}{\begin{tabular}[c]{@{}l@{}}$\star$ Having optically computation speed compared to conventional ANN.\end{tabular}}                                                                                                                                                       \\ \hline
\multicolumn{1}{l|}{AirComp}                             & \multicolumn{3}{l|}{\begin{tabular}[c]{@{}l@{}}Leveraging the wireless channel to realize\\simple addition of model parameters.\end{tabular}}                                                                                                                        & \multicolumn{3}{l}{\begin{tabular}[c]{@{}l@{}}$\star$ Leveraging programmable metasurfaces to realize\\controllable computations as EM waves propagate the SIM.\end{tabular}}                                                                                                                                                           \\ \hline\hline
\multicolumn{7}{c}{Comparison of SIM with other MIMO technologies}                           \\ \hline
\multicolumn{1}{l|}{MIMO technologies}                   & \multicolumn{1}{l|}{\begin{tabular}[c]{@{}l@{}}Number of\\ RF chains\end{tabular}} & \multicolumn{1}{l|}{\begin{tabular}[c]{@{}l@{}}ADC/DAC\\ resolutions\end{tabular}} & \multicolumn{1}{l|}{\begin{tabular}[c]{@{}l@{}}Hardware\\ cost\end{tabular}} & \multicolumn{1}{l|}{\begin{tabular}[c]{@{}l@{}}Energy\\ efficiency\end{tabular}} & \multicolumn{1}{l|}{\begin{tabular}[c]{@{}l@{}}Digital\\ precoder\end{tabular}} & \begin{tabular}[c]{@{}l@{}}Hardware implementation\end{tabular}                                                \\ \hline
\multicolumn{1}{l|}{Fully digital}                       & \multicolumn{1}{l|}{Large}                                                         & \multicolumn{1}{l|}{High}                                                     & \multicolumn{1}{l|}{High}                                          & \multicolumn{1}{l|}{Low}                                                         & \multicolumn{1}{l|}{\Checkmark}                                                        & \begin{tabular}[c]{@{}l@{}}Driving each antenna with an individual RF chain.\end{tabular}                      \\ \hline
\multicolumn{1}{l|}{Hybrid digital \& analog}            & \multicolumn{1}{l|}{Moderate}                                                      & \multicolumn{1}{l|}{Moderate}                                                 & \multicolumn{1}{l|}{Moderate}                                                    & \multicolumn{1}{l|}{Moderate}                                                    & \multicolumn{1}{l|}{\Checkmark}                                                        & \begin{tabular}[c]{@{}l@{}}Integrating phase shifters or metasurfaces with\\transceivers to reduce the number of RF chains.\end{tabular}          \\ \hline
\multicolumn{1}{l|}{Hybrid active \& passive} & \multicolumn{1}{l|}{Moderate}                                                      & \multicolumn{1}{l|}{Moderate}                                                      & \multicolumn{1}{l|}{Low}                                                & \multicolumn{1}{l|}{High}                                              & \multicolumn{1}{l|}{\Checkmark}                                                        & \begin{tabular}[c]{@{}l@{}}Deploying RISs in the environment to\\improve the wireless link quality.\end{tabular} \\ \hline
\multicolumn{1}{l|}{Proposed wave-based}                  & \multicolumn{1}{l|}{Small}                                                         & \multicolumn{1}{l|}{Low}                                                      & \multicolumn{1}{l|}{Very low}                                                & \multicolumn{1}{l|}{Very high}                                              & \multicolumn{1}{l|}{\XSolidBrush}                                                         & \begin{tabular}[c]{@{}l@{}}Leveraging multi-layer SIM to realize precoding\\and combining in the wave domain.\end{tabular}                \\ \hline
\end{tabular}\vspace{-0.5cm}
\end{table*}
To elaborate, Fig. \ref{fig_1} also contrasts an SIM-aided MIMO transceiver against the hybrid MIMO counterparts.
\begin{itemize}
\item \textbf{\emph{Hybrid Digital and Analog Beamforming:}} The popular eigenmode transmission is supported by digital beamforming at the baseband and by up-converting the superimposed signals to the RF bands, as shown at the top of Fig. \ref{fig_1}. Accordingly, the signals of different RF chains are steered toward the desired direction by performing analog beamforming through unit-modulus phase shifters across a larger number of transmit antennas \cite{CM_2014_Alkhateeb_MIMO}. At the receiver, a symmetric hybrid combining is adopted for recovering the data streams.
\item \textbf{\emph{Hybrid Active and Passive Beamforming:}} In the middle of Fig. \ref{fig_1}, the transmitter constructs multiple data streams via a digital beamformer by assigning a unique precoding vector to each data stream. Analog beamforming is performed passively in a far away located RIS for favorably reshaping the wireless propagation environments \cite{JSAC_2020_Di_Smart}, or near the transceivers to reduce the number of RF chains \cite{Access_2018_Zhou_Hardware}.
\item \textbf{\emph{Wave-Based Beamforming:}} The MIMO transceiver relying on a co-located SIM performs transmit precoding and receive combining in the wave domain, as depicted at the bottom of Fig. \ref{fig_1}. As such, multiple data streams are individually transmitted from preselected antennas \cite{JSAC_2022_An_Stacked}. At the receiver, each data stream is detected separately at the corresponding receive antenna.
\end{itemize}
The unique SIM architecture leads to the following distinctive beneficial features. Firstly, in contrast to the hybrid architecture whose performance is constrained by the analog beamformer relying on a set of unit-modulus phase shifters or metasurfaces, an SIM is capable of achieving full-precision digital beamforming by appropriately increasing the number of metasurface layers \cite{JSAC_2022_An_Stacked}. Secondly, an SIM that features a compact multilayer structure is strategically positioned in close proximity of the transceiver (e.g., acting as an antenna radome in Fig. \ref{fig_1}), which may not result in significant propagation loss as in the multi-hop systems and makes it more energy efficient than the family of holographic MIMO systems that rely on a substantial number of active components \cite{COM_2021_Dardari_Holographic}. Furthermore, an SIM aims to upgrade the MIMO transceiver design, and, therefore, it constitutes a complement of RIS for reshaping wireless propagation environments. For example, when the line-of-sight link at the bottom of Fig. \ref{fig_1} is blocked, one could also deploy an RIS into the wireless system to unleash the full potential offered by metasurfaces. Explicitly, we contrast our SIM and the relevant technologies and concepts in Table \ref{tab1}.

\subsection{Benefits of SIM}
We elaborate on the key advantages of the proposed SIM-aided MIMO transceiver in comparison to its conventional counterparts.
\subsubsection{\textbf{Improved Computational Efficiency}}
The most remarkable feature of an SIM is its novel wave-based signal processing capability, which has the following benefits.
\begin{itemize}
\item \emph{Ultrafast Computational Speed:} In contrast to conventional MIMO transceivers relying on digital signal processors, an SIM performs precoding and combining as the EM wave propagates through multiple metasurface layers. The calculation time of the WBF is determined by the thickness of the SIM \cite{NE_2022_Liu_A}. In an SIM having a thickness of $0.3$ m, for example, the precoding calculation can be completed within a nanosecond. As such, the SIM has great potential in ultra-reliable low-latency communication (URLLC) scenarios.
\item \emph{Parallel Computational Capability:} As a benefit of the intrinsic parallel propagation of the EM waves, an SIM is capable of simultaneously performing precoding and combining for all the data streams. The calculation time of the parallel forward propagation through the SIM is independent of the number of mathematical operations, which result in high processing complexity and delay in conventional RF-powered devices.
\item \emph{Reduced Computational Complexity:} Since MIMO precoding and combining are carried out in the wave domain, the proposed SIM-aided MIMO transceiver no longer requires conventional baseband signal processing techniques involving complex matrix inversion and decomposition \cite{CM_2014_Alkhateeb_MIMO}. Instead, only single-stream modulation and demodulation are required, which substantially reduces the signal processing complexity of the MIMO transceiver \cite{JSAC_2022_An_Stacked}.
\end{itemize}

\subsubsection{\textbf{Simplified Hardware Architecture}}
Since both MIMO precoding and combining operations are performed in the wave domain, the data stream of each user can be individually detected at the corresponding receive antenna \cite{JSAC_2022_An_Stacked}. Thus, the hardware architecture can be simplified from the following two aspects.
\begin{itemize}
\item \emph{Low-Resolution DAC/ADC:} Since the inter-stream interference is completely eliminated in the wave domain, it becomes feasible to use low-resolution power-efficient digital-to-analog converters (DAC) and analog-to-digital converters (ADC) for single-stream narrowband modulation and demodulation, respectively \cite{NE_2022_Liu_A}.
\item \emph{Reduced Number of RF Chains:} In contrast to mMIMOs relying on a large number of active components \cite{CM_2014_Alkhateeb_MIMO}, an SIM-aided MIMO transceiver is capable of approaching the full digital beamforming capability without driving each meta-atom by an individual active RF chain \cite{NE_2022_Liu_A}.
\end{itemize}
\begin{figure*}[!t]
\centering
\includegraphics[width=15cm]{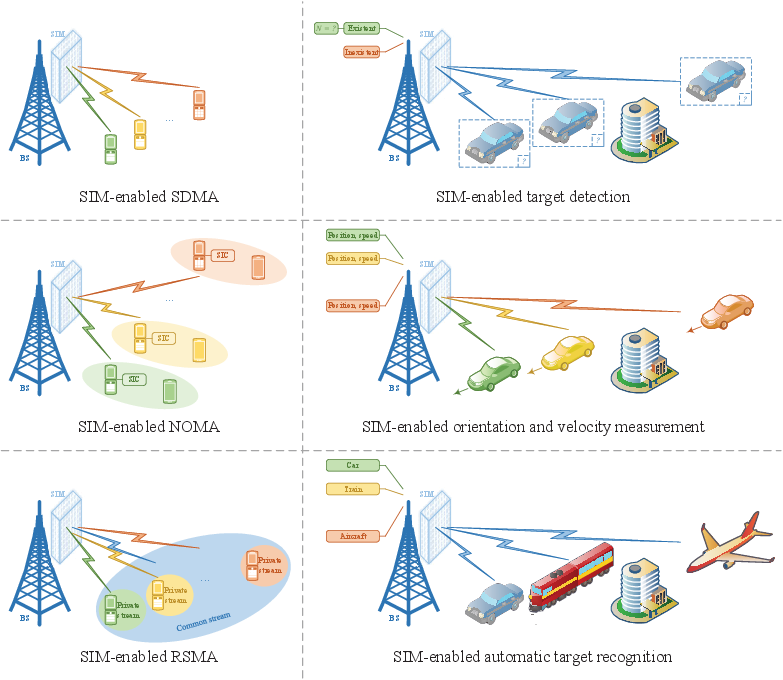}
\caption{Potential applications of SIM in typical sensing and communication scenarios.}\vspace{-0.5cm}
\label{fig_2}
\end{figure*}

\subsubsection{\textbf{Reduced Energy Consumption}}
On the one hand, the simplified hardware architecture consumes significantly reduced power for RF signal processing. On the other hand, an SIM attains array gains by utilizing the entire metasurfaces and by suppressing the co-channel interference in the wave domain, thus being capable of achieving a given performance target with less power for signal transmission \cite{JSAC_2022_An_Stacked}. As such, an SIM substantially reduces the overall energy consumption compared to conventional digital transceiver designs \cite{NE_2022_Liu_A}.

\section{Potential Applications and Open Challenges of SIM}
Next, we indicate the attractive applications and some open challenges of SIM in next-generation wireless networks.
\subsection{Potential Applications of SIM}
As illustrated in Fig. \ref{fig_2}, an SIM can be applied to perform interference cancellation for multiple access and to enable integrated sensing and communications \cite{JSAC_2022_An_Stacked, arXiv_2023_An_Stacked}.
\subsubsection{\textbf{Interference Suppression for Multiple Access}}
Multi-user interference is a crucial problem in existing cellular/WiFi networks \cite{TCOM_2022_An_Low}. Instead of implementing transmit beamforming in the baseband, an SIM is capable of performing WBF for effectively suppressing the multiuser interference, while substantially reducing the signal processing complexity and delay \cite{NE_2022_Liu_A}. Moreover, the WBF optimization can be incorporated into different multiple access strategies, e.g., space division multiple access (SDMA), non-orthogonal multiple access (NOMA), as well as rate-splitting multiple access (RSMA) \cite{CM_2022_de_Rate}. The tailored optimization procedures require future research. As shown on the left-hand side of Fig. \ref{fig_2}, by appropriately optimizing the SIM, one could suppress the inter-user interference, as the downlink signal propagates through the SIM. For the uplink, the data stream of each user can be directly detected at the corresponding antennas by applying low-resolution ADCs.

\subsubsection{\textbf{Integrated Sensing and Communications (ISAC)}}
Integrating radar sensing into 6G networks is envisioned to hold great promise in addressing a range of issues, including spectral congestion and energy efficiency. This has triggered an upsurge of research interests for ISAC, where these two functionalities share the same system modules and radio resources \cite{CST_2022_Zhang_Enabling}. The proposed SIM is expected to play a key role in typical ISAC scenarios. On the one hand, appropriate WBF designs efficiently suppress the mutual interference between these two signal components in a coexistence mode. Furthermore, an advanced SIM enables novel wave-based signal processing to retrieve information about the target of interest, such as its direction of arrival (DOA), location, and mobility. This is achieved automatically as the EM wave propagates through the SIM, as illustrated on the right-hand side of Fig. \ref{fig_2}. For example, by comparing the signal magnitude received at different probes, a well-trained SIM is capable of recognizing different targets \cite{Science_2018_Lin_All}. This means that the receiver hardware can be simplified by removing the power-thirsty ADCs.

Considering complex classification or regression tasks, one may introduce a non-linear module by making all the meta-atoms operate in their nonlinear range, thus further enhancing the inference capability of SIM \cite{NE_2022_Liu_A}. It is expected that the employment of SIM in ISAC systems may accomplish complex sensing tasks at a low processing delay, while the potential solutions in the context of SIM-aided ISAC deserve further exploration.

In addition to the aforementioned pair of scenarios, there are several other promising application scenarios for SIM in next-generation wireless networks, such as cell-free massive MIMO, simultaneous wireless information and power transfer, physical layer security, and index modulation. The employment of SIMs in these applications presents intriguing research directions. Furthermore, it becomes possible to perform mathematical operations such as the Fourier transform in the wave domain by employing SIMs. As a result, energy-efficient single-carrier RF devices are eminently suitable for generating multi-carrier waveforms through SIMs, which is especially beneficial for mmWave and terahertz hardware. In order to satisfy the stringent requirements for low latency and data privacy, federated learning has emerged for training machine learning (ML) models across network edges having limited computation, storage, and energy resources. To aggregate the distributed trained models, over-the-air computation (AirComp) approaches have shown promise by leveraging the signal superposition property of wireless multiple-access channels \cite{TWC_2020_Yang_Federated}. It is anticipated that SIMs are capable of performing more complex and controllable computations than the simple addition hence enhancing the performance of federating learning.

\subsection{Major Research Challenges and Opportunities}
Despite its compelling benefits, some practical challenges in implementing SIM-aided transceivers -- including the efficient SIM design, inter-layer transmission model, CSI acquisition, and WBF design -- call for further research.

\subsubsection{\textbf{Efficient SIM Design}}
\begin{figure}[!t]
\centering
\includegraphics[width=8cm]{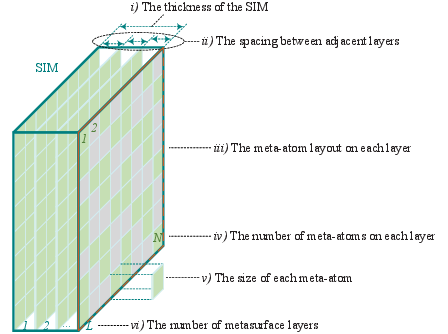}
\caption{Schematic of an SIM's major hardware parameters.}\vspace{-0.5cm}
\label{fig_2_0}
\end{figure}

As shown in Fig. \ref{fig_2_0}, an individual SIM hardware architecture has six main parameters, including \emph{i)} the thickness of the SIM; \emph{ii)} the spacing between adjacent layers; \emph{iii)} the meta-atom layout on each layer; \emph{iv)} the number of meta-atoms on each layer; \emph{v)} the size of each meta-atom, and finally \emph{vi)} the number of metasurface layers. By appropriately designing \emph{i), iv), v), vi)}, one could strike beneficial tradeoffs between the hardware cost, power consumption, transmission loss, as well as the computing capability of an SIM. Additionally, to enhance the computing capability of SIM, it may be feasible to beneficially leverage the EM coupling among closely packed meta-atoms and metasurfaces. This, however, requires an accurate model that characterizes the EM coupling within the SIM \cite{NE_2022_Liu_A}. Optimizing the response of the meta-atoms accounting for their EM coupling may have the potential of enhancing the SIM's computing capability even for a moderate number of metasurface layers. Finally, adjusting the spacing between adjacent metasurfaces provides an extra degree-of-freedom (DoF) for operating the SIM as a varifocal lens, which has the potential of focusing the energy on a target of interest with reduced leakage.

\subsubsection{\textbf{Inter-layer Transmission Model}}
The inter-layer transmission model is crucial for configuring an SIM. The transmission coefficients between adjacent metasurfaces can be approximately characterized by the Rayleigh-Sommerfeld diffraction theory of near-field propagation \cite{Science_2018_Lin_All}. However, the inter-layer transmission coefficients may deviate from the ones predicted by the transmission model due to the existence of practical hardware imperfections and inevitable fabrication shortcomings \cite{NE_2022_Liu_A}. As such, further research efforts are needed for experimentally validating the inter-layer transmission model. Additionally, the calibration of the transmission coefficients is imperative before practical SIM deployments. Specifically, by transmitting a known excitation signal and measuring the response at auxiliary probes, the transmission coefficients can be updated by employing the classic error back-propagation algorithm \cite{NE_2022_Liu_A}.

\subsubsection{\textbf{CSI Acquisition}}
The CSI acquisition in SIM-assisted communication systems is another critical challenge since an SIM operating without sensing modules cannot directly estimate the channels of users. Evidently, all the channels associated with $K$ users can be obtained by using at least $ \left \lceil KN/M \right \rceil$ pilot symbols, with $M$ denoting the number of RF chains at the BS. Meanwhile, the joint optimization of the SIM responses and the uplink pilots may further improve the channel estimation accuracy. To reduce both the complexity and pilot overhead, one could employ compressive sensing and deep learning methods by exploiting the channel sparsity in the beamspace \cite{Access_2018_Zhou_Hardware}. Additionally, codebook-based schemes are also promising options, which avoid the need for explicitly estimating the SIM-user channels \cite{TCOM_2022_An_Low}. Furthermore, the joint transmission coefficient calibration and channel estimation constitute an attractive research topic.
\subsubsection{\textbf{WBF Design}}
The WBF design of practical SIM-assisted communication systems is extremely challenging. On the one hand, although no digital beamforming is required, the WBF in general has to be jointly designed with the resource allocation at the BS (e.g., antenna selection and power allocation) to achieve the desired performance, which usually leads to intractable non-convex optimization problems that are challenging to solve optimally. On the other hand, practical SIMs are fabricated subject to various tuning constraints, e.g., discrete amplitude/phase shift levels of each meta-atom, which leads to an NP-hard problem formulation. A convenient approach is to first relax the non-convex constraints and then quantize the resultant solution to its nearest feasible values. The gradient descent method can then be employed for obtaining at least a locally optimal solution for the continuous phase-shift configuration. To improve the performance, a heuristic successive refinement technique might also be harnessed for optimizing the phase shift of each meta-atom one by one. Additionally, ML-based techniques, e.g., deep reinforcement learning, constitute a promising technique for optimizing WBF at a low complexity. Furthermore, the wideband WBF design relying on SIMs poses greater challenges. In contrast to conventional digital beamforming, which deals with different frequency sub-bands separately, SIMs have to address the resultant beam squint effect to fulfill the requirements of all frequency sub-bands.

\subsubsection{\textbf{Antenna Selection}}
Again, in contrast to conventional digital baseband signal processors assigning an individual beamforming vector to each information symbol, the new transceiver transmits each data stream directly from the corresponding antenna at the transmitter. This means that the transmitter has to select an appropriate number of antennas in advance to serve the same number of single-antenna users. The antenna selection problem is similar to that in conventional MIMO systems having a limited number of RF chains at the transmitter. Apart from considering the ordinary criterion, e.g., maximizing the equivalent signal-to-noise ratio at the receiver, one should also consider the SIM design. For example, selecting antennas having high spacing might be helpful for reducing inter-user interference. The joint antenna selection and WBF design constitutes a future research topic.

\subsubsection{\textbf{Performance Evaluation}}
The holistic performance evaluation of SIMs, including their power consumption and energy efficiency, requires conducting system-level experiments together with a numerically tractable power model, which however remains unavailable at the time of writing. According to the data measured in a practical SIM-aided system, one may characterize the power consumption by employing ML techniques, which can also implicitly account for the hardware imperfections, such as the penetration loss through the metasurfaces. Additionally, integrating micro-amplifiers into each meta-atom may provide a promising research direction for exploiting a large number of power-efficient amplifiers to replace the inefficient power amplifiers typically utilized at the transmitter \cite{NE_2022_Liu_A}. However, the fundamental trade-offs therein require further research. Moreover, a comprehensive assessment of the capability of an SIM of suppressing inter-cell interference in cellular networks, its compatibility with diverse duplexing modes and random access, and the impact of mutual coupling necessitate further exploration.

\section{Case Studies}
In this section, we provide numerical results for validating the wave-based signal processing capability of an SIM-aided efficient MIMO transceiver design.

\subsection{Multi-User Interference Cancellation}
\begin{figure}[!t]
\centering
	\subfloat[\label{fig_41}Simulation setup of an SIM-aided multiuser MISO wireless system.]{
		\includegraphics[width=7cm]{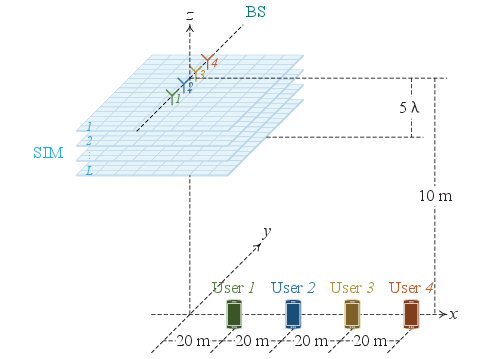}}\\
	\subfloat[\label{fig_42}Average sum-rate versus the number of metasurface layers.]{
		\includegraphics[width=7cm]{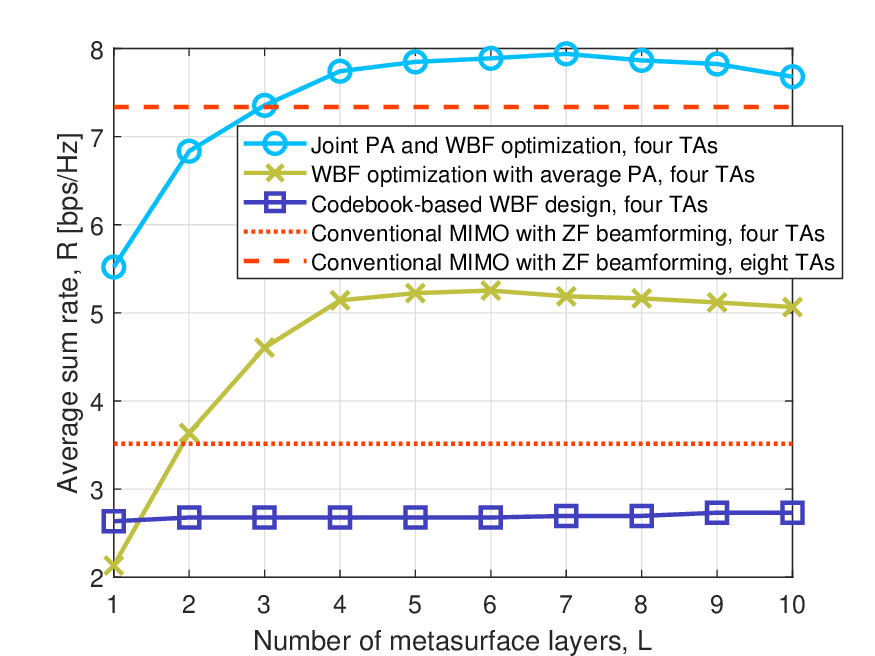}}
\caption{Multi-user interference cancellation using an SIM.}\vspace{-0.5cm}
\label{fig_4}
\end{figure}

We first characterize the capability of SIM for suppressing inter-user interference. As shown in Fig. \ref{fig_41}, we consider the case study where a four-antenna BS serves four single-antenna users. An SIM having $L$-layer metasurfaces, each consisting of $N$ programmable meta-atoms, is utilized for performing the transmit beamforming in the EM wave domain. All the metasurface layers in the SIM are placed parallel to the ground. The height of the BS is $10$ m, while the thickness of the SIM is $5\lambda$ with $\lambda$ representing the wavelength. Furthermore, we assume half-wavelength spacing between the adjacent antennas/meta-atoms of the BS and all the metasurfaces. The inter-layer transmission coefficients are determined by the Rayleigh-Sommerfeld diffraction formula \cite{Science_2018_Lin_All}. For simplicity, we consider fixed user locations with a spacing of $20$ m between the users (see Fig. \ref{fig_41}). The SIM-user channels are generated by applying a correlated Rayleigh fading model with a path loss exponent of $3.5$. The communication system operates at a carrier frequency of $28$ GHz. The transmit power of the BS is $20$ dBm, while the receiver noise power is $–100$ dBm for all the users. Moreover, we investigate the sum-rate maximization problem by considering the following four benchmark schemes: \emph{i)} Alternately optimizing the transmit power allocation (PA) using the iterative water filling and WBF through gradient descent algorithms \cite{JSAC_2022_An_Stacked}; \emph{ii)} Optimizing the WBF given the average PA solution; \emph{iii)} A codebook-based scheme with a randomly generated codebook of size $10LN$; \emph{iv)} The conventional zero-forcing (ZF) beamforming without using an SIM, and considering four and eight transmit antennas (TAs).

Fig. \ref{fig_42} shows the average sum-rate for different schemes versus the number of metasurface layers $L$, where we consider square-shaped metasurfaces having $N = 49$ meta-atoms on each layer. Jointly optimizing the PA and WBF provides a higher sum-rate for a small to moderate increase in $L$, benefiting from the capability of an SIM for effectively suppressing the inter-user interference in the EM wave domain \cite{ICC_2023_An_Stacked}. Nonetheless, the sum-rate peaks at $L = 7$ layers, which, compared to a single-layer SIM, provides a $44$\% increase in sum-rate, from $5.5$ bps/Hz to $7.9$ bps/Hz. Further increasing the number of layers leads to a denser metasurface arrangement and more severe signal loss, hence deteriorating the beamforming ability. As such, both the cost efficiency and signal processing capability of an SIM should be jointly considered in practical designs. Additionally, the joint optimization scheme outperforms all the benchmark schemes considered. Specifically, the average power allocation suffers a $2.5$ bps/Hz performance erosion in comparison to its optimal counterpart. Moreover, SIM-based systems operating without a digital beamformer rely heavily on the SIM configuration. However, a small codebook struggles to find the optimal phase shift from the entire solution space, thus providing negligible performance improvement upon increasing the number of metasurface layers. Finally, as compared to the conventional ZF beamforming operating without an SIM, the proposed SIM-aided transceiver attains more than $200$\% rate improvements due to the extra spatial gain achieved by the large aperture of the metasurface, despite its reduced signal processing complexity attained by removing the need for expensive digital beamforming.

\begin{figure}[!t]
\centering
	\subfloat[\label{fig_51}Simulation setup of SIM-aided DOA estimation.]{
		\includegraphics[width=7cm]{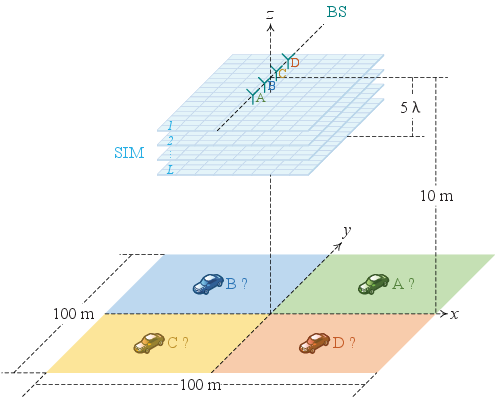}}\\
	\subfloat[\label{fig_52}Estimation accuracy rate versus the number of metasurface layers.]{
		\includegraphics[width=7cm]{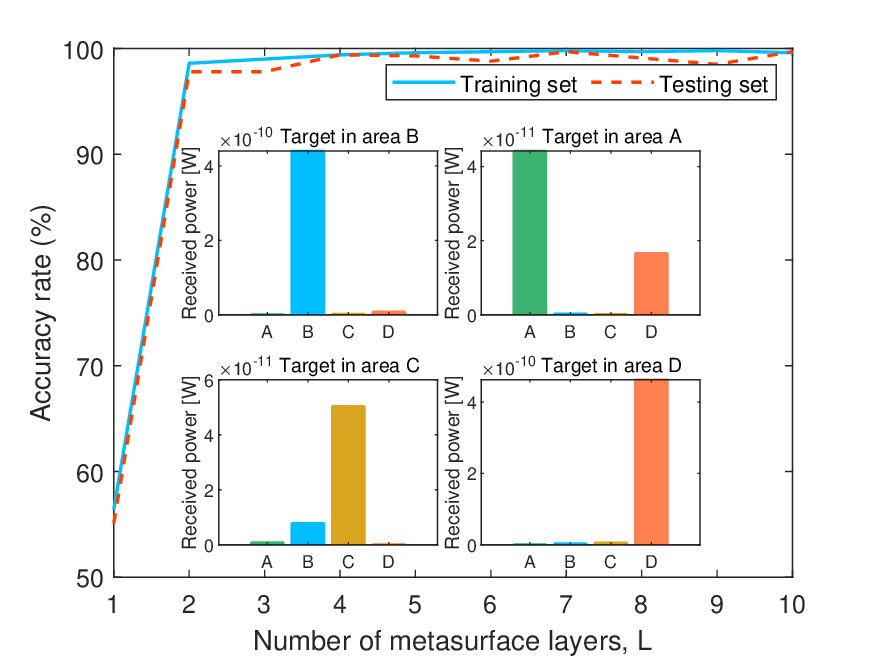}}
\caption{DOA estimation using an SIM.}\vspace{-0.5cm}
\label{fig_5}
\end{figure}
\subsection{DOA Estimation}
To demonstrate the versatility of an SIM, we next consider a typical radar sensing application by performing DOA estimation in the wave domain. The geographical layout is shown in Fig. \ref{fig_51}, where an $L$-layer SIM is integrated with a four-antenna BS to perform DOA estimation. Each metasurface is equipped with $100$ meta-atoms. For simplicity, we consider a rather rough estimation accuracy by dividing the square coverage area of $\left ( 100\times 100 \right )$ m$^2$ into four equal-size areas, i.e., {A, B, C, D} as shown in Fig. \ref{fig_51}. Each antenna at the BS corresponds to a potential area of the target. Specifically, the antenna having the maximum received energy indicates that the target is in the corresponding area. Note that measuring the amplitude of the received signals is equivalent to applying a nonlinear activation function that fits the SIM in performing nonlinear classifications. A reinforcement learning-based approach is employed for training the phase shifts of the SIM \cite{NE_2022_Liu_A}. The loss function used for evaluating the training process is the mean square error between a scaled version of the desired energy distribution and the energy received at the antenna array by inputting each of the training samples. After training with $1,000$ random samples, the SIM having appropriately designed phase shifts is utilized for testing another $100$ samples. The power of the signal transmitted by the target is set to $20$ dBm, while the noise power at each antenna is set to $-140$ dBm.

Fig. \ref{fig_52} shows the estimation accuracy as a function of the number of metasurface layers. An SIM having four metasurface layers is capable of correctly estimating the DOA of a target of interest with a detection success rate of $99.5$\% in terms of the training set. The SIM after training with $1,000$ samples achieves an accuracy of $99$\% in the testing set, which indicates that an appropriately designed SIM can directly determine the DOA of a target as the EM wave propagates through it, with nearly no extra time delay. For the sake of illustration, Fig. \ref{fig_52} also plots four specific examples by considering the target in different areas. Note that the EM waves radiated from the target would be focused on the corresponding receive antenna with the aid of a well-trained SIM. This simple toy example was used for confirming the versatile capability of SIM-aided MIMO transceivers. The specific design capable of performing various sensing tasks with high accuracy calls for future research.

\section{Summary and Conclusions}
The novel concept of SIM as a key enabler of smart MIMO transceiver design was proposed. Notably, the new MIMO transceiver architecture relying on a multilayer metasurface is capable of advanced signal processing, e.g., MIMO precoding and combining, directly in the wave domain. We elaborated on the available SIM hardware architectures and highlighted their potential in improving signal processing efficiency. Moreover, several promising SIM application scenarios have been identified. Nevertheless, the exploration of the emerging SIM concept is still in its infancy, with numerous unique challenges ahead, such as the efficient SIM design, the realistic inter-layer transmission model, the WBF optimization, and the CSI acquisition, to name but a few. Overall, we outlined a novel vision, where the advanced SIM concept is capable of fundamentally upgrading the MIMO transceiver architecture to enable ultra-fast and power-efficient signal processing in the wave domain, thus opening fertile directions for future research.

\bibliographystyle{IEEEtran}
\bibliography{ref.bib}

\begin{thebibliography}{10}
\providecommand{\url}[1]{#1}
\csname url@samestyle\endcsname
\providecommand{\newblock}{\relax}
\providecommand{\bibinfo}[2]{#2}
\providecommand{\BIBentrySTDinterwordspacing}{\spaceskip=0pt\relax}
\providecommand{\BIBentryALTinterwordstretchfactor}{4}
\providecommand{\BIBentryALTinterwordspacing}{\spaceskip=\fontdimen2\font plus
\BIBentryALTinterwordstretchfactor\fontdimen3\font minus
  \fontdimen4\font\relax}
\providecommand{\BIBforeignlanguage}[2]{{%
\expandafter\ifx\csname l@#1\endcsname\relax
\typeout{** WARNING: IEEEtran.bst: No hyphenation pattern has been}%
\typeout{** loaded for the language `#1'. Using the pattern for}%
\typeout{** the default language instead.}%
\else
\language=\csname l@#1\endcsname
\fi
#2}}
\providecommand{\BIBdecl}{\relax}
\BIBdecl

\bibitem{CM_2014_Alkhateeb_MIMO}
A.~Alkhateeb, \emph{et al.}, ``{MIMO} precoding and combining solutions for
  millimeter-wave systems,'' \emph{IEEE Commun. Mag.}, vol.~52, no.~12, pp.
  122--131, Dec. 2014.

\bibitem{CST_2022_Zhang_Enabling}
J.~A. Zhang, \emph{et al.}, ``Enabling joint communication and radar sensing in
  mobile networks—a survey,'' \emph{IEEE Commun. Surs. Tuts.}, vol.~24,
  no.~1, pp. 306--345, 1st Quarter 2022.

\bibitem{JSAC_2020_Di_Smart}
M.~Di~Renzo, \emph{et al.}, ``Smart radio environments empowered by
  reconfigurable intelligent surfaces: How it works, state of research, and the
  road ahead,'' \emph{IEEE J. Sel. Areas Commun.}, vol.~38, no.~11, pp.
  2450--2525, Nov. 2020.

\bibitem{TCOM_2022_An_Low}
J.~An, C.~Xu, L.~Gan, and L.~Hanzo, ``Low-complexity channel estimation and
  passive beamforming for {RIS}-assisted {MIMO} systems relying on discrete
  phase shifts,'' \emph{IEEE Trans. Commun.}, vol.~70, no.~2, pp. 1245--1260,
  Feb. 2022.

\bibitem{TCOM_2022_Liu_Compact}
K.~Liu, Z.~Zhang, L.~Dai, and L.~Hanzo, ``Compact user-specific reconfigurable
  intelligent surfaces for uplink transmission,'' \emph{IEEE Trans. Commun.},
  vol.~70, no.~1, pp. 680--692, Jan. 2022.

\bibitem{COM_2021_Dardari_Holographic}
D.~Dardari and N.~Decarli, ``Holographic communication using intelligent
  surfaces,'' \emph{IEEE Commun. Mag.}, vol.~59, no.~6, pp. 35--41, Jun. 2021.

\bibitem{NE_2022_Liu_A}
C.~Liu, \emph{et al.}, ``A programmable diffractive deep neural network based
  on a digital-coding metasurface array,'' \emph{Nature Electro.}, vol.~5,
  no.~2, pp. 113--122, Feb. 2022.

\bibitem{Science_2018_Lin_All}
X.~Lin, \emph{et al.}, ``All-optical machine learning using diffractive deep
  neural networks,'' \emph{Science}, vol. 361, no. 6406, pp. 1004--1008, Jul.
  2018.

\bibitem{arXiv_2023_An_Stacked}
J.~An, C.~Yuen, M.~Di~Renzo, M.~Debbah, H.~V. Poor, and L.~Hanzo, ``Stacked
  intelligent metasurface performs a {2D DFT} in the wave domain for {DOA}
  estimation,'' \emph{arXiv preprint arXiv:2310.09861}, 2023.

\bibitem{ChemPhysMater_2022_Yu_Electromagnetic}
S.~Yu, J.~Cheng, Z.~Li, W.~Liu, H.~Cheng, J.~Tian, and S.~Chen,
  ``Electromagnetic wave manipulation based on few-layer metasurfaces and
  polyatomic metasurfaces,'' \emph{ChemPhysMater}, vol.~1, no.~1, pp. 6--16,
  Jan. 2022.

\bibitem{JSAC_2022_An_Stacked}
J.~An \emph{et~al.}, ``Stacked intelligent metasurfaces for efficient
  holographic {MIMO} communications in {6G},'' \emph{IEEE J. Sel. Areas
  Commun.}, vol.~41, no.~8, pp. 2380--2396, Aug. 2023.

\bibitem{ICC_2023_An_Stacked}
J.~An, M.~Di~Renzo, M.~Debbah, and C.~Yuen, ``Stacked intelligent metasurfaces
  for multiuser beamforming in the wave domain,'' in \emph{Proc. IEEE Int.
  Conf. Commun. (ICC)}, Rome, Italy, May 2023, pp. 1--6.

\bibitem{Access_2018_Zhou_Hardware}
Z.~Zhou, N.~Ge, Z.~Wang, and S.~Chen, ``Hardware-efficient hybrid precoding for
  millimeter wave systems with multi-feed reflectarrays,'' \emph{IEEE Access},
  vol.~6, no.~1, pp. 6795--6806, Jan. 2018.

\bibitem{CM_2022_de_Rate}
A.~S. de~Sena, \emph{et al.}, ``Rate-splitting multiple access and its
  interplay with intelligent reflecting surfaces,'' \emph{IEEE Commun. Mag.},
  vol.~60, no.~7, pp. 52--57, Jul. 2022.

\bibitem{TWC_2020_Yang_Federated}
K.~Yang, T.~Jiang, Y.~Shi, and Z.~Ding, ``Federated learning via over-the-air
  computation,'' \emph{IEEE Trans. Wireless Commun.}, vol.~19, no.~3, pp.
  2022--2035, Mar. 2020.

\end{thebibliography}
\section*{Biographies}
\vspace{-1cm}
\begin{IEEEbiographynophoto}{Jiancheng An}
is currently a Research Fellow with the Engineering Product Development Pillar, Singapore University of Technology and Design, Singapore.
\end{IEEEbiographynophoto}
\vspace{-1cm}
\begin{IEEEbiographynophoto}{Chau Yuen}
is currently an Associate Professor at the Singapore University of Technology and Design. He is an Editor of IEEE Transactions on Communications and IEEE Transactions on Vehicular Technology.
\end{IEEEbiographynophoto}
\vspace{-1cm}
\begin{IEEEbiographynophoto}{Chao Xu}
is currently a Senior Research Fellow working with the Next Generation Wireless Research Group, University of Southampton.
\end{IEEEbiographynophoto}
\vspace{-1cm}
\begin{IEEEbiographynophoto}{Hongbin Li}
is currently the Charles and Rosanna Batchelor Memorial Chair Professor with the Department of Electrical and Computer Engineering, Stevens Institute of Technology, Hoboken, NJ, USA.
\end{IEEEbiographynophoto}
\vspace{-1cm}
\begin{IEEEbiographynophoto}{Derrick Wing Kwan Ng}
is currently a Scientia Associate Professor with the University of New South Wales, Sydney, Kensington, NSW, Australia. He is currently the Editor of IEEE Transactions on Communications, IEEE Transactions on Wireless Communications, and the Area Editor of IEEE Open Journal of the Communications Society.
\end{IEEEbiographynophoto}
\vspace{-1cm}
\begin{IEEEbiographynophoto}{Marco Di Renzo}
is a CNRS Research Director (Professor) and the Head of the Intelligent Physical Communications in the Laboratory of Signals and Systems at Paris-Saclay University CNRS and CentraleSupelec. He serves as the Editor-in-Chief of IEEE Communications Letters.
\end{IEEEbiographynophoto}
\vspace{-1cm}
\begin{IEEEbiographynophoto}{M\'erouane Debbah}
is currently the Chief Researcher with the Technology Innovation Institute, Abu Dhabi, UAE. He was the recipient of more than 25 best paper awards from major IEEE conferences and journals.
\end{IEEEbiographynophoto}
\vspace{-1cm}
\begin{IEEEbiographynophoto}{Lajos Hanzo}
received Honorary Doctorates from the Technical University of Budapest and Edinburgh University. He is a Foreign Member of the Hungarian Science-Academy, Fellow of the Royal Academy of Engineering (FREng), of the IET, of EURASIP and holds the IEEE Eric Sumner Technical Field Award.
\end{IEEEbiographynophoto}
\vfill
\end{document}